\documentclass{elsart}
\usepackage{graphicx,amssymb}
\journal{Physics Letters B}
\begin{document}
\begin{frontmatter}


\title{Schizophrenic active neutrinos and exotic sterile neutrinos }
\author{A. C. B. Machado\thanksref{ana}},
\ead{ana@ift.unesp.br}
\author{V. Pleitez\thanksref{vp}}
\address{Instituto de F\'\i sica Te\'orica,
Universidade Estadual Paulista\\ Rua Pamplona 145,
01405-900 - S\~ao Paulo, SP, Brazil
}
\ead{vicente@ift.unesp.br}
\thanks[ana]{Supported by CNPq}
\thanks[vp]{Partially supported by CNPq under the process
302102/2008-6}



\begin{abstract}
We implement a schizophrenic scenario for the active neutrinos in a model in which there are also
exotic right-handed neutrinos making a model with a local $U(1)_{_{B-L}}$ anomaly free.
Two of right-handed neutrinos carry $B-L=-4$ while the third one carries $B-L=5$. Unlike the non-exotic
version of the model, in which all right-handed neutrinos carry the same $B-L=-1$ charge,
in this case the neutrinos have their own scalar sector and no hierarchy in the Yukawa
coupling in the Dirac mass term is necessary.
\end{abstract}
\end{frontmatter}


It is well known that $B-L$ is an automatic anomalous global symmetry of the degrees of
freedom of the standard model~\cite{thooft}. In order to make this a local symmetry
is necessary to add right-handed neutrinos (that are sterile with respect to the standard model interactions).
For instance, three of them with the same lepton number than the left-handed (active) neutrinos. It was
shown in \cite{blsm} that, working within a $SU(2)_L\otimes U(1)_{Y^\prime}\otimes U(1)_{_{B-L}}$ electroweak
model, there exist other solutions in which the number of right-handed neutrinos are not necessarily three or,
they have other $B-L$ charge assignments.
In particular, in the case of three right-handed neutrinos there is a solution to the anomalies cancelation in which
two right-handed neutrinos have $B-L=-4$ and the third one has $B-L=5$.

On the other hand, recently, it was shown that it is possible to have \textit{schizophrenic} neutrinos~\cite{moha10}:
the neutrinos of all flavor are part Dirac and part Majorana, in particular one of the neutrino
mass eigenstates is, at the tree level, Dirac  whereas the other two are Majorana.

Here we shall show that in models with exotic right-handed neutrinos we can implement a scenario
in which the active neutrinos are of the schizophrenic type. The mechanism can of course be implemented in
the non-exotic version the model also considered in Ref.~\cite{blsm} i.e., that in which all right-handed neutrinos
carry the same $B-L=-1$ assignment. This version is in fact almost similar to the model
considered in \cite{moha10}. In this case the mass eigenstates having the Dirac mass must have a extremely tiny
Yukawa coupling. This is because the VEV appearing in the Dirac mass term is the same that also gives
mass to the $u$-quark. The exotic version on the other hand predict that neutrinos have their own scalar sector
and the VEVs are not necessarily large thus, avoiding the hierarchy in the Yukawa coupling mentioned above.

The quantum number of  leptons and scalars of the model are shown in Table~\ref{table1}. As in \cite{moha10} we impose
$S_3$ symmetry which permute the three families among themselves~\cite{s3}, that is they are in a triplet of $S_3$:
($\textbf{3}=L_e,L_\mu,L_\tau$), this is a reducible representation since
$\textbf{3}=\textbf{2}_L+\textbf{1}_L$. We can define in the lepton sector~\cite{japa}
\begin{eqnarray}
&&\textbf{1}_L\equiv L_2=\frac{1}{\sqrt3}(L_e+L_\mu+L_\tau)\nonumber \\ &&
\textbf{2}_L\equiv D_L= (L_1,L_3)=\left(\frac{1}{\sqrt6}(2L_e-L_\mu-L_\tau),\frac{1}{\sqrt2}(L_\mu-L_\tau)\right),
\nonumber \\&&
\textbf{1}_{\mu R}\equiv n_{\mu R},
\quad \textbf{2}_{e\tau R}\equiv N_R=(n_{eR},n_{\tau R}).
\label{21}
\end{eqnarray}

The scalar sector consists of two additional doublets, in relation to the standard model which scalar is denoted by $\Phi_{_{SM}}$,
with weak hypercharge $Y=-1$ i.e., $\Phi_i=(\varphi^0_i\;\varphi^-_i)^T$ that are singlets of $S_3$ and three singlets ($Y=0$),
forming a doublet of $S_3$, $\Delta=(\phi_1,\phi_2)$ and a singlet $\phi_3$. See Table~\ref{table1}.
We will also impose the discrete $Z_3$ symmetry under which
$L_2,\Phi_1, n_{\mu R}$, and $\Delta$ transform as $\omega$ and, $D_L,N_R$, $\Phi_2$ and $\phi_3$ transform as $\omega^2$,
the other fields transform trivially under $Z_3$.
With these fields we obtain the following Yukawa interactions in the lepton sector (quarks are
assumed to be singlets under $S_3$) that is invariant under the gauge symmetries and $S_3\otimes Z_3$ are
\begin{eqnarray}
-\mathcal{L}^\nu_{\textrm{yukawa}}&=&
h_1\bar{L}_2\Phi_1n_{\mu R}+ y_1[\bar{D}_L\otimes N_R]_1\Phi_2+
\frac{y_2}{\Lambda}[\overline{(N_R)^c}\otimes \Delta]_1[N_R\otimes \Delta]_1
\nonumber \\ &+&y_3\phi_3\,[\overline{(N_{ R})^c}\otimes N_{R}]_1
+H.c.
\label{yuka1}
\end{eqnarray}
We impose that $y_1v_2\ll y_2u_{1,2}/\Lambda\ll y_3u_3$ in order to Majorana
masses dominate in the $(n_{e R},n_{\tau R})$ sector (the notation is $\langle\varphi^0_{1,2}\rangle=v_{1,2}/\sqrt{2}$,
$\langle\phi_{1,2,3}\rangle=u_{1,2,3}/\sqrt{2}$). The main contribution to the Majorana masses for the singlets
$n_{eR}$ and $n_{\tau R}$ comes from the $y_3$ interactions
but, they have different Majorana masses due to the interaction $y_2$.
Under those conditions, the interaction proportional to $y_1$ is relevant mainly to generate
the vertex $D_LN_R\Phi_2$.

After integrating out $n_{e R}$ and $n_{\tau R}$ we obtain the effective
interactions~\cite{weinberg}
\begin{eqnarray}
-\mathcal{L}^{\textrm{eff}}_{\textrm{yukawa}}&=&h_1\bar{L}_2\Phi_1n_{\mu R}+
 \frac{h^2_2}{m_{n_e}}\,(L_1\Phi_2)^2+
\frac{h^2_3}{m_{n_\tau}}\,(L_3 \Phi_2)^2
+H.c.,
\label{yuka2}
\end{eqnarray}
where the mixing angles in the $(n_{eR},n_{\tau R})$ sector have been absorbed in $h_2$ and $h_3$.
Thus, we have the Yukawa interactions in which one neutrino, $\nu_2$, has at the tree level a Dirac mass term
particle $m^D_2=h_1v_1$.  On the other hand, the Majorana mass matrix generated by effective interactions
(\ref{yuka2}) that, at the leading order, is (in the $\nu_e,\nu_\mu,\nu_\tau$ basis)
\begin{equation}
M^\nu_M=\frac{h^2_2v^2_2}{m_{n_e}}\left(\begin{array}{ccc}
\frac{2}{3}&-\frac{1}{3}&-\frac{1}{3}\\
-\frac{1}{3}&\frac{1}{6}+\frac{h^2_3}{h^2_2}\frac{m_{n_e}}{2m_{n_\tau}}&\frac{1}{6}-\frac{h^2_3}{h^2_2}\frac{m_{n_e}}{2m_{n_\tau}}\\
-\frac{1}{3}&\frac{1}{6}-\frac{h^2_3}{h^2_2}\frac{m_{n_e}}{2m_{n_\tau}}&\frac{1}{6}+\frac{h^2_3}{h^2_2}\frac{m_{n_e}}{2m_{n_\tau}}
\end{array}\right),
\label{massa}
\end{equation}
which is a consequence of the $S_3$ symmetry~\cite{moha06}.  This matrix is diagonalized at the leading order by a
tri-bi-maximal matrix~\cite{harrison}:
\begin{equation}
U=\left(\begin{array}{ccc}
\sqrt{\frac{2}{3}}&-\frac{1}{\sqrt3}&0\\
-\frac{1}{\sqrt6}&\frac{1}{\sqrt3}&-\frac{1}{\sqrt2}\\
-\frac{1}{\sqrt6}&\frac{1}{\sqrt3}&-\frac{1}{\sqrt2}
\end{array}\right),
\label{tribi}
\end{equation}
that is the PMNS matrix if the charge leptons are assumed already in a diagonal
basis.

The eigenvalues in the active neutrino sector from Eq.~(\ref{massa}) are
$m^M_1=h^2_2v^2_2/m_{n_e}$, $m^M_2=0$, and $m^M_3=h^2_3v^2_2/m_{n_\tau}$. The Majorana
massive neutrinos are $\nu_1$ and $\nu_3$ while $\nu_2$ has only a Dirac at the tree level $m^D_2$ as we see above. If
$m^D_2\sim0.1$ eV and the Majorana masses $\sim10^{-2}$ eV we have the right neutrino
mass square differences observed in oscillation experiments~\cite{nus}. The inverted neutrino mass hierarchy is a prediction of the
model: $m^D_1\equiv m_1>m^M_1\equiv m_2\gg m^M_3\equiv m_3$ if $m_{n_\tau}\gg m_{n_e}$ and $h_2\sim h_3\stackrel{<}{\sim} O(1)$. As we will show below radiative corrections
give to $\nu_2$ a small Majorana mass and this neutrino is a pseudo-Dirac one~\cite{pseudo}.
Solar neutrino data constrain Majorana masses to be $10^{-9}$ eV if all Dirac masses are assumed to be larger than
the Majorana masses~\cite{gouvea}, however, it does not apply to the present case since we are in a situation in which only
one of the neutrinos is pseudo-Dirac.

\begin{table}
\begin{eqnarray*}
\begin{array}{|c||r|r|r|r|r|r|}\hline
\phantom{u_L} & I_3 & I &  Y^\prime & B-L & Y &Z_3\\ \hline\hline
\nu_{aL}     & 1/2 & 1/2  & 0 & -1 &-1 &\omega\\ \hline
l_{aL} & -1/2 & 1/2 &0 & -1 &-1 &\omega \\ \hline
e_{aR}     & 0 & 0   & -1& -1 &-2 & 1\\ \hline
n_{(e,\tau)R} & 0 & 0   & 4 & -4 & 0 &\omega^2 \\ \hline
n_{\mu R}        & 0 & 0   & -5 & 5 & 0 &\omega\\ \hline
\Phi_{_{SM}} & 1/2 & 1/2  & 0 & 0 &+1  &1\\ \hline
\Phi_1 & 1/2 & 1/2  & 5  & -6  & -1 & \omega\\ \hline
\Phi_2 & 1/2 & 1/2  & -4  & 3  & -1 & \omega^2 \\ \hline
\phi_{1,2} & 0 &0  &-4 &4 &0 & \omega\\ \hline
\phi_3 & 0 & 0  &  -8  & 8 &0 & \omega^2\\ \hline
\phi_x &0 & 0 &  6 &-6 &0  & \omega  \\ \hline
\phi_y & 0 & 0  &  -3 & 3 &0 &\omega^2 \\ \hline
\end{array}
\end{eqnarray*}
\caption{Quantum number assignment in the model. Quarks have the quantum number as usual. The singlets $\phi_{x,y}$ ($Y=0$)
do not interact with any fermion. See the text.}
\label{table1}
\end{table}

After the breaking of the electroweak and $S_3$ symmetries contributions to
the neutrino masses induced by one loop radiative corrections. For instance, (\ref{yuka2}) implies interactions like
$(h_2/\Lambda)(\overline{(\nu_{aL}^c)}\varphi^{0*}_2+ \overline{(l_{aL})^c}\varphi^+_2)(\nu_{bL}\varphi^{0*}_2+
l_{bL}\varphi^+_2),\;a,b=e,\mu,\tau$, the vertices from these interactions are $\sim h_2v_2/\Lambda$.
On the other hand, the scalar sector of the model has three $SU(2)$ doublets one which give mass to quarks and charge
leptons $\Phi_{SM}=(\varphi^+_{SM}\, \varphi^0_{SM})^T$, and the two exotic
doublets carrying B-L charge $\Phi_{1,2}$. Hence, there exist in the scalar potential terms like
$\lambda (\Phi^\dagger_2\Phi_{SM})(\Phi_{SM}^\dagger\Phi_2)$, implying a mixing in the mass matrix among the charged
scalars, i.e., $\lambda(\varphi^+_{SM}\varphi^{0*}_2\varphi^-_{SM}\varphi^0_2+ H.c.)$, and the mixing of $\varphi^-_{SM}$ with $\varphi^-_2$ is $\sim \lambda v^2_2$
(we are working in the flavor basis). When these corrections are taken into account the corrections to the Majorana
mass matrix (\ref{massa}) that are given by (up to logarithmic terms)~\cite{babu}
\begin{equation}
m^M_{ab}\approx \xi\,\frac{\lambda h^2_2V^2_S}{8\pi^2}\,\frac{v^3_2}{m^2_H\Lambda}
 \,\frac{U_{ba} m_{l_a}m_{l_b}}{v_{SM}},
\label{babu}
\end{equation}
(there is no summation over repeated indices),  where $\Lambda\sim m_{n_e},m_{n_\tau}$,
$\xi=2/3$ if $a=b$ and $\xi=-1/3$ if $a\not=b$; $U_{ab}$
is the tribimaximal mixing matrix (\ref{tribi});
$V^2_S$ denotes the mixing angles in the charged scalar sector,
$m_H$ denotes a typical value for the masses in the charged scalars sector,
$m_{l_a}$ is the mass of the charged leptons $a=e,\mu,\tau$ and $v_{SM}$ is the value
of the SM Higgs scalar.  Assuming all dimensionless parameters in (\ref{babu}) are $\sim O(1)$,
the Dirac mass $M^D_2$ gain corrections smaller than $10^{-4}$ depending if $v^3_2/m^2_H\Lambda\sim0.0246$.
The tribimaximal mixing matrix has to be considered as a leading order of the PMNS matrix, correction that turns it more realistic
might arise if the charged lepton mass matrix is almost diagonal, as that in Ref.~\cite{dim5},
hence this may induce a small $\theta_{13}$ angle. Recent global $\theta_{13}$ analysis implies that $\sin\theta_{13}=0.009^{+0.013}_{-0.007}$~\cite{kamland}.

With all the scalar fields shown in Table~\ref{table1}, the scalar potential invariant under the gauge symmetry of the SM and $A_4\otimes Z_3$ is
\begin{eqnarray}
&&V_{B - L} = \mu^2_{_{SM}} |\Phi_{SM}|^2 + \mu_1^2 |\Phi_1|^2 +  \mu_2^2 |\Phi_2|^2 + \mu_3^2 \left[\Delta^{*} \Delta\right]_1 + \mu_4^2 |\phi_3|^2
\\ \nonumber &+&
\mu^2_x\vert\phi_x\vert^2+\mu^2_y\vert\phi_y\vert^2+\lambda_x\vert\phi^2_x\vert^4+\lambda_y\vert\phi_y\vert^4+
\lambda_{_{SM}} (\Phi_{_{SM}}^{\dag} \Phi_{_{SM}})^2  + \lambda_{1} (\Phi_1^{\dag} \Phi_1)^2 + \lambda_{2} (\Phi_2^{\dag} \Phi_2)^2
 \\ \nonumber &+&\lambda_3 |\Phi_1|^2|\Phi_2|^2+\lambda_4\vert\Phi_{_{SM}}\vert^2\vert \Phi_1\vert^2 +
\lambda_5\vert\Phi_{_{SM}}\vert^2\vert \Phi_2\vert^2  + \lambda_6 (\Phi_1^{\dag} \Phi_2) (\Phi_2^{\dag} \Phi_1)+
\lambda_7(\Phi^\dagger_1\Phi_{_{SM}})(\Phi^\dagger_{_{SM}}\Phi_1 )\\ \nonumber &+&
\lambda_8 (\Phi^\dagger_2\Phi_{_{SM}})(\Phi^\dagger_{_{SM}}\Phi_2 )+
 \lambda_9 (\left[ \Delta^{*} \Delta\right]_1)^2 +  \lambda_{10} \left[\left[ \Delta^{*} \Delta\right]_{1^{\prime}} \left[ \Delta^{*} \Delta  \right]_{1^{\prime}}\right]_1+ \lambda_{11}  |\phi_3|^4 + \lambda_{12} |\Phi_{_{SM}}|^2 \left[ \Delta^{*} \Delta\right]_1
\\ \nonumber &+&\lambda_{13} |\Phi_{_{SM}}|^2  |\phi_3|^2 + \lambda_{14} |\Phi_1|^2 \left[ \Delta^{*} \Delta\right]_1 + \lambda_{15}|\Phi_2|^2 \left[ \Delta^{*} \Delta\right]_1+ \lambda_{16} |\Phi_1|^2  |\phi_3|^2+
\lambda_{17} |\Phi_2|^2  |\phi_3|^2 \\ \nonumber &+& \lambda_{18}  \left[ \Delta^{*} \Delta\right]_1  |\phi_3|^2 +(\lambda_{xy}\Phi^\dagger_1\Phi_2\phi_x\phi_y+\kappa[\Delta\Delta]_1\phi^*_3+\kappa_x\Phi^T_1\epsilon\Phi_{_{SM}}\phi_x+\kappa_y
\Phi^T_2\epsilon\Phi_{_{SM}}\phi_y+H.c.),
\label{potencial}
\end{eqnarray}
where we must take into account that $\textbf{2}\otimes\textbf{2}=\textbf{1}+\textbf{1}^\prime+ \textbf{2}$, $\textbf{1}^\prime\otimes\textbf{1}^\prime=\textbf{1}$~\cite{japa}. 
The Higgs potential above without the singlets $\phi_{x,y}$, has three extra global $U(1)$ symmetries.
If these symmetries are not extended to the Yukawa interactions there are three pseudo-Goldstone bosons~\cite{sw}. 
In the model of \cite{blsm}, in which this model is based, there are two $U(1)$ extra global symmetries.
In that case the pseudo Goldstone are eliminated by introducing one singlet singlets~\cite{msv}. Thus, for this reason we introduce 
in the present model two extra singlet denoted by $\phi_x$ and $\phi_y$ (see Table~\ref{table1}).

The conditions $\frac{\partial V_{B-L}}{\partial \phi_i}\large\vert_{_{\phi_i=V_i}}=0$,
$\phi_i=V_i/\sqrt{2},\;\; V_i=v_{_{SM}},v_{1,2},u_{1,2,3},v_v,v_y$ imply [after shifting the neutral component $\eta^0=(1/\sqrt2)(V_i+Re\, \eta^0+iIm\,\eta^0)$]
\begin{eqnarray}
\label{droga}
 && v_{_{SM}} \left[2\mu^2_{_{SM}} + 2 \lambda_{_{SM}} v_{_{SM}}^2  +\lambda_7v^2_1+\lambda_8 v^2_2+\lambda_{12} (u_1^2 + u_2^2 ) +\lambda_{13} u_3^2 \right]+\kappa_xv_1v_x+\kappa_2v_2v_y=0,
 \\ \nonumber  &&
 v_1 \left[2\mu_1^2 +  2\lambda_1 v_1^2+ ( \lambda_3 +\lambda_6 ) v_2^2 +\lambda_{7}v^2_{_{SM}}+ \lambda_{14} (u_1^2 + u_2^2)+\lambda_{16}u^2_3 \right]+\kappa_xv_{_{SM}}v_x+\lambda_{19}v_2v_xv_y=0,
 \\ \nonumber &&
 v_2 \left[2\mu_2^2 +  2\lambda_2 v_2^2 +( \lambda_3 + \lambda_6) v_1^2+ \lambda_8v^2_{_{SM}} +\lambda_{15} (u_1^2 + u_2^2 ) +\lambda_{17} u_3^2 \right]+\kappa_yv_{_{SM}}v_y+\lambda_{19}v_1v_xv_y=0,
 \\ \nonumber &&
u_1 [2\mu_3^2 +  2\lambda_9 (u_1^2 + u_2^2 ) +  2\lambda_{10} (u_1^2 - u_2^2 ) + \lambda_{12} v_{_{SM}}^2 + \lambda_{14} v_1^2 + \lambda_{15} v_2^2+ \lambda_{18} u_3^2+\sqrt{2}\kappa u_3]=0,
\\ \nonumber &&
u_2 [2\mu_3^2 +  2\lambda_9 (u_1^2 + u_2^2 ) +  2\lambda_{10} (u_1^2 - u_2^2 )  + \lambda_{12} v_{_{SM}}^2 + \lambda_{14} v_1^2 + \lambda_{15} v_2^2+ \lambda_{18} u_3^2+\sqrt{2}\kappa u_3 ]=0,
\\ \nonumber &&
u_3 \left[2\mu_4^2 + 2\lambda_{11} u_3^2  + \lambda_{13} v_{_{SM}}^2 + \lambda_{16} v_1^2 + \lambda_{17} v_2^2
+ \lambda_{18} (u_1^2 + u_2^2 )\right]+\sqrt{2}\kappa(u^2_1+u^2_2)=0,\\ \nonumber &&
v_x[2\mu^2_x+2\lambda_xv^2_x+\kappa_xv_1v_{_{SM}}+\lambda_{xy}v_1v_2v_y]=0,
\\ \nonumber &&
v_y[2\mu^2_y+2\lambda_yv^2_y+\kappa_yv_{_{SM}}v_2+\lambda_{xy}v_1v_2v_x]=0
\end{eqnarray}
From equations (\ref{droga}) we see that $u_1=u_2\equiv u$ (an $S_2$ symmetry remains unbroken). Solutions with $\mu^2_1>0,\mu^2_2>0$ are then possible. Since, if $\kappa_{1,2}=0$ the symmetries of the model increase these parameters may be naturally smaller that the electroweak scale,
the same for the VEVs $v_{x,y}$ if they are not the main responsible for the breaking of the $B-L$ symmetry (and for the masses of the $Z^\prime$ vector boson). Hence, there also solutions with $\mu^2_{1,2}\gg \vert\kappa_xv_x\vert,\vert\kappa_y v_y\vert,v_xv_y$. Then, we have
$2\lambda_{_{SM}}v^2_{_{SM}}\approx -2\mu^2_{_{SM}}-2\lambda_{12}u^2-\lambda_{13}u^2_3$, and
\begin{eqnarray}
v_1\approx \frac{\kappa_xv_x}{2\mu^2_1+2\lambda_{14}u^2+\lambda_{16}u^2_3}\,v_{_{SM}},\;
v_2\approx \frac{\kappa_yv_y}{2\mu^2_2+2\lambda_{15}u^2+\lambda_{17}u^2_3}\,v_{_{SM}},
\label{boa}
\end{eqnarray}
Therefore, we see that it is easy to obtain solutions to the above equations having the following hierarchy: $u\sim u_3\gg v_{_{SM}}\gg v_1> v_2$, independently of the values of the dimensionless $\lambda$s. Hence $v_{1,2}$ appearing in the neutrino Yukawa effective interactions (\ref{yuka2}) may be smaller than the others and the hierarchy in the Yukawa couplings appears in the VEVs values which numerical values are hidden under the
mechanism of spontaneous symmetry breaking. In spite the low value of $v_{1,2}$, the respective neutral fields are heavy since they
have masses $\sim\mu^2_{1,2}$~\cite{ma}.

The model has in the scalar sector three $SU(2)$ doublets, $\Phi_{_{SM}}$ ($Y=1$) and $\Phi_{1,2}$ ($Y=-1$) and five
scalar singlets ($Y=0$), $\phi_{1,2}$, $\phi_3$ nd $\phi_{x,y}$. All of them but $\Phi_{_{SM}}$ carry $B-L$ charge, while $\phi_{1,2,3}$
couple in the flavor basis only to the right-handed neutrinos, $\phi_{x,y}$ do not couple with
any fermion of the model. The VEV of the doublets $\Phi_{1,2}$ my be smaller than $v_{_{SM}}\sim174$ GeV. 
This implies that the Yukawa couplings may take natural values, $\stackrel{<}{\sim}O(1)$,
and the Majorana masses $m_{n_e}$ and $m_{n_\tau}$ do not need to be very large as well.
What possibility is the most interesting will depend on the following:
i) if there exist a combination of the neutral scalar components, say $\xi$, incorporated into a single flat direction
and, for this reason, driven the inflation; ii) the scalar singlets and/or the heavy right-handed neutrinos
can be dark matter candidates; iii) the decay of the scalar singlets and/or the heavy right-handed neutrinos can generate the
observed asymmetry through a soft leptogenesis mechanism. We are working on these possibilities in the supersymmetric
version of the model. Finally, we must stress that this sort of models has a new neutral vector boson which mass is related
to the scalar singlet VEVs and if it is of the order of TeVs, the boson may be discover at the LHC and study with more precision
at the ILC~\cite{zprime}.



\begin{thebibliography}{99}
\bibitem{thooft} G. t' Hooft, Phys. Rev. Lett. \textbf{37} (1976) 8.
\bibitem{blsm} J. C. Montero and V. Pleitez,  Phys. Lett. \textbf{B675} (2009) 64.
\bibitem{moha10} R. Allahverdi, B. Dutta, and R. N. Mohapatra, Phys. Lett. \textbf{B695} (2011) 181, arXiv:1008.1232.
\bibitem{s3} P.F. Harrison, W.G. Scott, Phys. Lett. B 333 (1994) 471, 
E. Derman, D.R.T. Jones, Phys. Lett. \textbf{B70} (1977) 449;
S. L. Adler, Phys. Rev. D \textbf{59} (1999) 015012.
\bibitem{japa} H. Ishimori \textit{et al}., arXiv:1003.3552v2.
\bibitem{weinberg} S. Weinberg, Phys. Rev. Lett. \textbf{43} (1979) 1566; F. Wilczek and
A. Zee, Phys. Rev. Lett. \textbf{43} (1979) 1571; E. Ma, phys. Rev. Lett. \textbf{81} (1998) 1171.
\bibitem{moha06} R. N. Mohapatra, S. Nasri, and H-B. Yu, Phys. Lett. \textbf{B639} (2006) 318.
\bibitem{harrison} P. F. Harrison, D. Perkins, and W. G. Scott, Phys. Lett. \textbf{530} (2002) 167.
\bibitem{nus} B. Aharmin \textit{et al}., Phys. Rev. Lett. \textbf{101} (2008) 111301;
M.H. Ahn,  \textit{et al}., Phys. Rev. D \textbf{74}, 072003 (2006); Y. Ashie \textit{et al}., Phys.
Rev. Lett. \textbf{93} (2004) 101801.
\bibitem{pseudo} L. Wolfenstein, Nucl. Phys. \textbf{B 186} (1981) 147.
\bibitem{gouvea} A. de Gouvea and W-C. Huang, Phys. Rev. D \textbf{80} (2008) 073007.
\bibitem{babu}
K. S. Babu and V. S. Mathur, Phys. Rev. D \textbf{38} (1988) 3550.
\bibitem{dim5} A.C.B. Machado, V. Pleitez,  Phys. Lett. \textbf{B674} (2009) 223.

\bibitem{kamland} A. Gando \textit{et al}. (KamLAND Collaboration), arXiv:1009.4771v2.
\bibitem{sw} S. Weinberg, Phys. Rev. Lett. \textbf{29}  (1972) 1698.
\bibitem{msv} J. C. Montero and B. S\'anchez-Vega, arXiv:1102.0321.
\bibitem{ma} E. Ma and U. Sarkar, Phys. Rev. Lett. \textbf{80} (1998) 5716.
\bibitem{zprime} E. C. F. S. Fortes, J. C. Montero, and V. Pleitez, Phys. Rev. D \textbf{82} (2010) 114007,
arXiv:1005.2991.

\end{thebibliography}
\end{document}